\documentclass[a4paper,12pt]{iopart}
\usepackage{iopams}
\usepackage[dvips]{graphicx}
\usepackage{longtable}
\usepackage{multirow}
\usepackage{color}
\long\def\symbolfootnote[#1]#2{\begingroup%
\def\thefootnote{\fnsymbol{footnote}}\footnote[#1]{#2}\endgroup} 
\newcommand {\nc} {\newcommand}
\nc {\beq} {\begin{eqnarray}}
\nc {\eeq} {\end{eqnarray}}
\nc {\eeqn} [1] {\label{#1} \end{eqnarray}}
\nc {\eoln} [1] {\label{#1} \\}
\nc {\eol} {\nonumber \\}
\nc {\la} {\mbox{$\langle$}}
\nc {\ra} {\mbox{$\rangle$}}
\nc {\cL} {\mbox{${\cal L}$}}
\nc {\dem} {\mbox{$\frac{1}{2}$}}
\nc {\ve} [1] {\mbox{\boldmath $#1$}}
\nc {\arrow} [2] {\mbox{$\mathop{\rightarrow}
\limits_{#1 \rightarrow #2}$}}
\begin{document}
\title{Quadrupole transitions in the bound rotational-vibrational spectrum 
of the hydrogen molecular ion}
\author{Horacio Olivares~Pil\'on and Daniel Baye}  
\address{Physique Quantique C.P. 165/82, and 
Physique Nucl\'eaire Th\'eorique et Physique Math\'ematique, C.P.\ 229, 
Universit\'e Libre de Bruxelles (ULB), B 1050 Brussels, Belgium}
\eads{\mailto{dbaye@ulb.ac.be}}
\begin{abstract}
The three-body Schr\"odinger equation of the H$_2^+$ hydrogen molecular ion with Coulomb potentials 
is solved in perimetric coordinates using the Lagrange-mesh method. 
The Lagrange-mesh method is an approximate variational calculation 
with variational accuracy and the simplicity of a calculation on a mesh. 
Energies and wave functions of up to four of the lowest vibrational bound or quasibound states 
for total orbital momenta from 0 to 40 are calculated. 
The obtained energies have an accuracy varying from about 13 digits for the lowest vibrational state 
to at least 9 digits for the third vibrational excited state. 
With the corresponding wave functions, a simple calculation using the associated Gauss quadrature 
provides accurate quadrupole transition probabilities per time unit 
between those states over the whole rotational bands. 
Extensive results are presented with six significant figures. 
\end{abstract}
\pacs{31.15.Ag, 31.15.Ac, 02.70.Hm, 02.70.Jn}
\submitto{\jpb}

\centerline{\today}
\maketitle
\section{Introduction}
\label{sec:intro}
The H$_2^+$ bound rotational-vibrational spectrum possesses about 420 bound states 
corresponding to the $\Sigma_g$ electronic configuration as well as about 60 narrow quasibound levels. 
For this simple three-body system, the Schr\"odinger equation involving Coulomb potentials 
can not be solved exactly but it is possible to reach a high accuracy for both energies 
and wave functions. 
A calculation of the energies of most rotational-vibrational bound states and some 
quasi-bound levels has been performed in 1993 with 11-digit accuracy by Moss \cite{Mo93}. 
He also calculated energies for the few extended bound $\Sigma_u$ states. 
Since then, several calculations have reached a higher accuracy, but always for a 
limited number of bound states. 
The energy of the ground state, with total orbital momentum $L = 0$ and positive parity, 
has been improved in a number of papers \cite{Ko00,HBG00,BF02,YZL03,CD04,LWZ07,HNN09} 
to reach an accuracy around or beyond 30 digits \cite{LWZ07,HNN09}. 
The energy of the first $L = 0$ excited state has also been improved \cite{CD04,HNN09}. 
The lowest $L = 1$ vibrational state has been considered by several authors 
\cite{YZL03,CD04,LWZ07,HNN09}. 
The $L = 0$ and $1$ excited vibrational states have been studied by Hilico \etal \cite{HBG00} 
(see also \cite{CLD03}). 
Energies of the full ground-state rotational band were determined with 12-digit accuracy 
by Hesse and Baye \cite{HB03} and energies up to $L = 12$ with higher accuracy 
were obtained by Yan and Zhang \cite{YZ04}. 
Some of the quoted works also provide accurate information on mean radii \cite{HB03,HNN09} 
or quadrupole moments \cite{HB03}. 

In opposition to this large number of accurate studies, very few studies concern 
the transition probabilities between the rotational-vibrational bound states of H$_2^+$. 
There are several reasons for this. 
As the electric dipole transitions are forbidden by the symmetry the two protons,
the more complicated electric quadrupole transitions are the dominant mode of decay. 
Moreover, few existing studies provide the necessary wave functions. 
Since the pioneering work of Bates and Poots \cite{BP53}, a systematic study 
of all transitions for states up to $L=20$ within the Born-Oppenheimer approximation 
has been published with two significant figures \cite{PDP83}. 
Here we present accurate E2 transition probabilities without Born-Oppenheimer 
approximation. 
They are obtained from three-body wave functions calculated 
with the Lagrange-mesh method in perimetric coordinates \cite{HB99,HB01,HB03}, 
with which the calculation is particularly simple and very precise. 

The Lagrange-mesh method is an approximate variational calculation 
using a basis of Lagrange functions and the associated Gauss quadrature. 
It has the high accuracy of a variational approximation and the simplicity of a calculation 
on a mesh. 
Lagrange functions are $N$ orthonormal infinitely differentiable functions 
that vanish at all points of this mesh, except one. 
Used as a variational basis in a quantum-mechanical calculation, 
the Lagrange functions lead to a simple approximation 
when matrix elements are calculated with the associated Gauss quadrature. 
The variational equations take the form of mesh equations with a diagonal representation 
of the potential only depending on values of this potential at the mesh points 
\cite{BH86,Ba06}. 
The most striking property of the Lagrange-mesh method is that, in spite of its simplicity, 
the obtained energies and wave functions can be as accurate with the Gauss quadrature 
approximation as in the original variational method 
with an exact calculation of the matrix elements \cite{BHV02,Ba06}. 
The accuracy of the lowest energies exceeds by far the accuracy of the Gauss quadrature 
for the individual matrix elements. 
The Lagrange-mesh method allows very accurate calculations not only in simple 
quantum-mechanical problems but also in various more complicated applications 
in atomic \cite{BVH08,BHV08,BS10}, molecular \cite{HB99,HB01,HB03,VB06,OBT10}
and nuclear \cite{DDB03,DTB06,DBD10} physics. 

In the H$_2^+$ case, the Lagrange-mesh method is applied in perimetric coordinates, 
i.e. three angles describing the orientation of the plane containing the particles 
and three linear combinations of the distances between them \cite{Pe58}. 
The dependence on the three Euler angles is treated analytically \cite{HB01,HB03}. 
The three perimetric coordinates vary from zero to infinity and can easily 
be discretized on a three-dimensional Lagrange-Laguerre mesh \cite{HB99}. 
An additional advantage is that the resulting matrix is rather sparse. 
The Lagrange-mesh method also provides analytical approximations for the wave functions 
that lead to very simple expressions for a number of matrix elements 
when used with the corresponding Gauss-Laguerre quadrature. 

In section \ref{sec:lmtp}, the basic expressions for the transition probabilities 
are recalled. 
The E1 and E2 operators are expressed in perimetric coordinates. 
Some definitions about Lagrange functions lead to Lagrange-mesh expressions 
for the transition matrix elements. 
In section \ref{sec:res}, energies are given for the lowest four vibrational 
states over the full rotational bands and E2 transition probabilities are tabulated. 
Concluding remarks are presented in section \ref{sec:conc}. 
\section{Lagrange-mesh calculation of transition probabilities}
\label{sec:lmtp}
\subsection{Oscillator strength and transition probability per time unit}
\label{sec:ostp}
The dimensionless oscillator strength for an electric transition of multipolarity $\lambda$ 
between an initial state $i$ and a final state $f$ is defined as \cite{So72,So79}
\beq
f_{i\rightarrow f}^{(\lambda)} = \frac{m_e c}{\hbar}\frac{(2\lambda+1)(\lambda+1)}{[(2\lambda+1)!!]^2\lambda} 
k^{2\lambda-1} \frac{S_{if}^{(\lambda)}}{2J_i+1} 
\eeqn{2.1}
where $m_e$ is the electron mass, 
\beq
k = \frac{|E_f-E_i|}{\hbar c}
\eeqn{2.2}
is the photon wavenumber and 
\beq
S_{if}^{(\lambda)} = S_{fi}^{(\lambda)} 
= \sum_{M_iM_f\mu} |\la \gamma_i J_i M_i| O^{(\lambda)}_\mu | \gamma_f J_f M_f \ra |^2 
= |\la \gamma_i J_i || O^{(\lambda)} || \gamma_f J_f \ra |^2,
\eeqn{2.3}
where $J_{i,f}$ is a total angular momentum, $M_{i,f}$ is its projection, $\gamma_{i,f}$ represents 
the other quantum numbers and the reduced matrix element is defined according to \cite{Ed57}. 
The transition irreducible tensor operator is given in units of $e$ by 
\beq
O^{(\lambda)}_\mu = \sum_i Z_i r_i^{\prime\lambda} C^{(\lambda)}_\mu (\Omega'_i)
\eeqn{2.4}
where $\ve{r}'_i = \ve{r} - \ve{R}_{\rm c.m.}$ is the relative coordinate of particle $i$ 
with respect to the center of mass, $Z_i$ is its charge in units of $e$ 
and $C^{(\lambda)}_\mu (\Omega) = \sqrt{4\pi/(2\lambda+1)} Y^{(\lambda)}_\mu (\Omega)$. 
Notice that the charge unit $e$ is included in the coefficient 
in $f_{i\rightarrow f}^{(\lambda)}$.
One has 
\beq
f_{f\rightarrow i}^{(\lambda)} = \frac{2J_i+1}{2J_f+1} f_{i\rightarrow f}^{(\lambda)}.
\eeqn{2.5}
If atomic units are used, the oscillator strength reads 
\beq
f_{i\rightarrow f}^{(\lambda)} = \frac{(2\lambda+1)(\lambda+1)}{[(2\lambda+1)!!]^2\lambda} 
\alpha^{2\lambda-2} \left(E_f - E_i\right)^{2\lambda-1} \frac{S_{if}^{(\lambda)}}{2J_i+1}
\eeqn{2.6}
where $\alpha$ is the fine-structure constant. 

The transition probability per time unit for $E_f < E_i$ is given in atomic units 
(the atomic unit of time is $a_0/\alpha c \approx 2.4188843 \times 10^{-17}$ s) by 
\beq
W_{i\rightarrow f}^{(\lambda)} 
= \frac{2(\lambda+1)(2\lambda+1)}{\lambda[(2\lambda+1)!!]^2} \alpha^{2\lambda+1} 
(E_i - E_f)^{2\lambda+1} \frac{S_{if}^{(\lambda)}}{2J_i+1}
\eeqn{2.8}
where all quantities are in a.u. 
For any multipolarity $\lambda$, 
the transition probability is related to the oscillator strength by 
\beq
W_{i\rightarrow f}^{(\lambda)} = 2\alpha^3 (E_i-E_f)^2 f_{i\rightarrow f}^{(\lambda)}.
\eeqn{2.10}
\subsection{Dipole and quadrupole operators in perimetric coordinates}
\label{sec:dqo}
After elimination of the centre-of-mass motion, the Hamiltonian 
depends on the two Jacobi coordinates $\ve{R}$ of proton 2 with respect to proton 1 
and $\ve{r}$ of the electron with respect to the centre of mass of both protons. 
These coordinates can be expressed as three Euler angles ($\psi, \theta, \phi$) 
defining the orientation of the triangle formed by the three particles, 
and three internal coordinates describing the shape of this triangle. 
The $\theta$ and $\psi$ angles correspond to the angular spherical coordinates of vector $\ve{R} = (R,\theta,\psi)$ 
and the $\phi$ angle is the angular cylindrical coordinate of vector $\ve{r} = (\rho,\zeta,\phi)$ 
in the relative frame where the $z$-axis is moved along $\ve{R}$ by $\psi$ and $\theta$ rotations \cite{Fe84}. 
For the internal degrees of freedom we use the perimetric coordinates ($x,y,z$) 
defined as linear combinations of interparticle distances \cite{Pe58}, 
\beq
\begin{array}{rcl}
x &=& R-r_{e2}+r_{e1}, \\
y &=& R+r_{e2}-r_{e1}, \\
z &=& -R+r_{e2}+r_{e1},
\end{array}
\eeqn{3.1}
where $r_{e1}$ and $r_{e2}$ are the distances between the electron and protons 1 and 2, respectively. 
The domains of variation of these six variables are 
$[0,2\pi]$ for $\psi$ and $\phi$, $[0,\pi]$ for $\theta$ and $[0,\infty[$ for $x, y$ and $z$. 

In the body-fixed frame, the radial component of $\ve{R}$ and the polar and axial 
components of $\ve{r}$ are expressed in perimetric coordinates \eref{3.1} as \cite{HB01,HB03}
\beq
R &=& \frac{x+y}{2},
\eoln{3.7}
\rho &=& \sqrt{\frac{xyz(x+y+z)}{(x+y)^2}},
\eoln{3.8}
\zeta &=& \frac{(x-y)(2z+x+y)}{4(x+y)}.
\eeqn{3.9}

For H$_2^+$, the dipole tensor operator reads in Jacobi coordinates, 
\beq
d^{(1)}_\mu = - \left(1 + \frac{m_e}{M}\right) r^{(1)}_\mu
\eeqn{3.2}
where $M=2m_p+m_e$ is the total mass of the molecular ion. 
This operator changes sign under space reflection (odd operator). 
It is invariant under the permutation of the protons.
The tensor components of $\ve{r}$ can be written 
as a function of the Euler angles $(\psi,\theta,\phi)$ as 
\beq
r^{(1)}_\mu = \zeta D^{1}_{\mu 0}(\psi,\theta,\phi) 
+ \frac{\rho}{\sqrt{2}} [D^{1}_{\mu 1}(\psi,\theta,\phi) - D^{1}_{\mu -1}(\psi,\theta,\phi)].
\eeqn{3.3}
In both terms, the Wigner matrices change sign under space reflection \cite{HB01} 
while $\zeta$ and $\rho$ remain unchanged. 
With respect to proton exchange \cite{HB01}, $\zeta$ and $D^{1}_{\mu 0}$ are both odd 
while $\rho$ and $D^{1}_{\mu 1} - D^{1}_{\mu -1}$ are both even. 

In the Jacobi coordinate system, the E2 tensor operator reads 
\beq
Q^{(2)}_\mu = \sqrt{\frac{3}{2}} \left\{ \frac{1}{2} [R^{(1)} \otimes R^{(1)}]^{(2)}_\mu 
- \gamma [r^{(1)} \otimes r^{(1)}]^{(2)}_\mu \right\}
\eeqn{3.6}
where 
\beq
\gamma = 1 - \frac{2m_e}{M} - \frac{m_e^2}{M^2}.
\eeqn{3.5}
In perimetric coordinates, it becomes 
\beq \fl
Q^{(2)}_\mu = \frac{1}{2} [R^2 - \gamma (2\zeta^2 - \rho^2)] D^{2}_{\mu 0}(\psi,\theta,\phi) 
- \sqrt{\frac{3}{2}} \gamma \zeta \rho [D^{2}_{\mu 1}(\psi,\theta,\phi) - D^{2}_{\mu -1}(\psi,\theta,\phi)]
\eol
- \sqrt{\frac{3}{8}} \gamma \rho^2 [D^{2}_{\mu 2}(\psi,\theta,\phi) + D^{2}_{\mu -2}(\psi,\theta,\phi)].
\eeqn{3.4}
Operator $Q^{(2)}_\mu$ is even with respect to parity and to permutation. 
\subsection{Transition matrix elements}
\label{sec:wfme}
The three-body Hamiltonian that we consider involves Coulomb forces between the particles 
but no spin-dependent forces. 
Hence the total orbital momentum $L$ and parity $\pi$ are good quantum numbers 
corresponding to constants of motion. 
The wave functions with orbital momentum $L$ and parity $\pi$ are expanded as \cite{HB03} 
\beq
\Psi^{(L^{\pi})\sigma}_M = \sum_{K=0}^L {\cal D}_{MK}^{L\pi}(\psi,\theta,\phi) \Phi_K^{(L^{\pi})\sigma}(x,y,z). 
\eeqn{4.1}
In practice, the sum can be truncated at some value $K_{\rm max}$. 
The normalized angular functions ${\cal D}_{MK}^{L\pi}(\psi,\theta,\phi)$ are defined for $K \ge 0$ by
\beq
{\cal D}_{MK}^{L\pi}(\psi,\theta,\phi) &=& \frac{\sqrt{2L+1}}{4\pi} \left(1+\delta_{K0}\right)^{-1/2}
\left[D_{MK}^L(\psi,\theta,\phi) \right.\nonumber\\
&& \left.+ \pi (-1)^{L+K} D_{M\;-K}^L(\psi,\theta,\phi)\right]
\eeqn{4.2}
where $D_{MK}^L(\psi,\theta,\phi)$ represents a Wigner matrix element. 
They have parity $\pi$ and change as $\pi(-1)^K$ under permutation of the protons. 
Hence $\Phi_K^{(L^{\pi})\sigma}$ is symmetric for $(-1)^K = \sigma\pi$ and antisymmetric for 
$(-1)^K = -\sigma\pi$, when $x$ and $y$ are exchanged. 
Most bound states belong to the $\Sigma_g$ band where $K=0$ dominates and $\sigma$ is equal to $\pi$. 

Since the symmetry of the proton spin part is $(-1)^{S+1}$ where $S$ is the total spin of the protons, 
physical states (i.e.\ states antisymmetric with respect to the exchange of the protons) have $\sigma = (-1)^S$. 
In the $\Sigma_g$ rovibrational band, states have natural parity, $\pi = \sigma = (-1)^L$. 
Even-$L$ states have positive parity and $\sigma = +1$. 
The protons are thus in a singlet state and the total intrinsic spin of the molecule is $1/2$. 
Odd-$L$ states have negative parity and $\sigma = -1$. 
The protons are in a triplet state and the total intrinsic spin of the molecule is $1/2$ or $3/2$. 
E1 transitions are forbidden because of different proton symmetries when $\Delta L = 1$. 
However, they are possible from the three weakly bound states of the $\Sigma_u$ band 
where $\pi = \sigma = (-1)^{L+1}$. 
E2 transitions are possible within the $\Sigma_g$ rovibrational band for $\Delta L = 0$, $\pm 2$. 
Such states have the same parity and symmetry.

Using the property 
\beq
\la D^{L'}_{M'K'} | D^\lambda_{\mu\kappa} | D^{L}_{MK} \ra = \frac{8\pi^2}{2L'+1}(L\lambda M\mu|L'M') (L\lambda K\kappa|L'K'),
\eeqn{4.3}
one can write for $\kappa \ge 0$, 
\beq
F_{K K';\kappa}^{L L';\lambda} 
& = & \la {\cal D}_{M'K'}^{L'\pi'} | D^\lambda_{\mu\kappa} + (-1)^\kappa D^\lambda_{\mu-\kappa}| {\cal D}_{MK}^{L\pi} \ra 
\eol
& = & \frac{(2L+1)^{1/2}}{[(2L'+1)(1+\delta_{K'0}) (1+\delta_{K0})]^{1/2}} (L\lambda M\mu|L'M')
\eol 
& \times & \{ (L \lambda K \kappa|L' K') + (-1)^\kappa (L \lambda K -\!\!\kappa|L' K') 
\eol
& + & \pi'(-1)^{L'+K'} [(L \lambda K \kappa|L' -\!\!K') + (-1)^\kappa (L \lambda K -\!\!\kappa|L' -\!\!K') ] \}. 
\eeqn{4.4}

The reduced matrix elements of the operators $Q^{(2)}_\mu$ between initial and final states 
$\Psi_{M_i}^{i(L_i^{\pi_i})\sigma_i}$ and $\Psi_{M_f}^{f(L_f^{\pi_f})\sigma_f}$ 
can be written as 
\beq
\la \Psi^{f(L_f^{\pi_f})\sigma_f} || Q^{(2)} || \Psi^{i(L_i^{\pi_i})\sigma_i} \ra 
= \sum_{K_i K_f} \sum_{\kappa = 0}^2 (1+\delta_{\kappa 0})^{-1} 
F_{K_i K_f;\kappa}^{L_i L_f;2} A_{K_i K_f;\kappa}^{L_i L_f}
\eeqn{4.5}
where the perimetric matrix elements 
\beq
A_{K_i K_f;\kappa}^{L_i L_f} = \la \Phi_{K_f}^{f(L_f^{\pi_f})\sigma_f} | \mathcal{A_\kappa} | \Phi_{K_i}^{i(L_i^{\pi_i})\sigma_i} \ra
\eeqn{4.6}
are calculated by integration over the perimetric coordinates with the volume element 
$(x+y)(y+z)(z+x)dxdydz$ and, from \eref{3.4}, 
\beq
\mathcal{A}_0 = \frac{1}{2} [R^2 - \gamma (2\zeta^2 - \rho^2)], \ \ 
\mathcal{A}_1 = - \sqrt{\frac{3}{2}} \gamma \zeta \rho, \ \ 
\mathcal{A}_2 = - \sqrt{\frac{3}{8}} \gamma \rho^2,
\eeqn{4.7}
with $R$, $\rho$ and $\zeta$ replaced by their expressions \eref{3.7}-\eref{3.9}.
With \eref{2.3} and \eref{4.5}, the oscillator strength is given explicitly 
for transitions between natural-parity states $\pi_{i,f} = (-1)^{L_{i,f}}$ by 
\beq 
S^{(2)}_{if} = (2L_i+1) \bigg| \sum_{K_i K_f} \left\{ C^{L_i2L_f}_{K_i0K_f} A_{K_i K_f;0}^{L_i L_f} \right. 
\eol 
+ \left[ C^{L_i2L_f}_{K_i1K_f}  (1+\delta_{K_i 0})^{1/2} - C^{L_i2L_f}_{K_i-1K_f}  (1+\delta_{K_f 0})^{1/2} \right] 
A_{K_i K_f;1}^{L_i L_f} 
\eol \fl
\left. + \left[ C^{L_i2L_f}_{K_i2K_f}  (1+\delta_{K_i 0})^{1/2} + C^{L_i2L_f}_{K_i-2K_f}  (1+\delta_{K_f 0})^{1/2} 
- C^{L_i2L_f}_{2-11} \delta_{K_i1} \delta_{K_f1} \right] A_{K_i K_f;2}^{L_i L_f} \right\} \bigg|^2\,
\eeqn{4.8}
where the Clebsch-Gordan coefficients are written as $C^{L_i \lambda L_f}_{K_i\mu K_f} = (L_i \lambda K_i\mu|L_fK_f)$ 
and the sums over $K_i$ and $K_f$ are truncated at $K_{\rm max}$ in practice. 
The remaining calculation of the matrix elements $A_{K_i K_f;\kappa}^{L_i L_f}$ is particularly simple 
within the Lagrange-mesh method as shown in the next subsection. 
\subsection{Lagrange-mesh method}
\label{sec:lmm}
The three-dimensional Lagrange functions $F^K_{ijk}(x,y,z)$ are infinitely differentiable functions defined by
\beq
F^K_{ijk}(x,y,z)= {\cal N}_{Kijk}^{-1/2} {\cal R}_K(x,y,z) f_i^{N_x}(x/h_x)f_j^{N_y}(y/h_y) f_k^{N_z}(z/h_z).
\eeqn{5.1}
The one-dimensional Lagrange-Laguerre functions $f_i^N$ are given by
\beq
f_i^N(u)=(-1)^i u_i^{1/2} \frac{L_N(u)}{u-u_i} e^{-u/2}
\eeqn{5.2}
where $L_N(u)$ is the Laguerre polynomial of degree $N$ and $u_i$ is one of its zeros, i.e.\ $L_N(u_i)=0$. 
They vanish at all $u_j$ with $j \ne i$. 
Basis \eref{5.1} is exactly equivalent to the set of functions 
${\cal R}_K(x,y,z) L_{n_x} (x/h_x) L_{n_y} (y/h_y) L_{n_z} (z/h_z) \exp[-(x/2h_x)-(y/2h_y)-(z/2h_z)]$ 
with $n_{x,y,z} = 0$ to $N_{x,y,z}-1$. 
The mesh points $(h_x u_p,h_y v_q,h_z w_r)$ correspond to the zeros $u_p$, $v_q$, $w_r$ of Laguerre polynomials 
of respective degrees $N_x$, $N_y$, $N_z$. 
Three scale parameters $h_x, h_y, h_z$ are introduced in \eref{5.1} in order to fit the different meshes 
to the size of the actual physical problem. 
The function ${\cal R}_K(x,y,z)$ is a regularization factor introduced 
because of the presence of singularities in the Hamiltonian operator when $L$ differs from zero \cite{HB01,HB03}. 
It is equal to 1 when $K=0$ and to $\sqrt{xyz(x+y+z)}$ otherwise. 
The normalization factor ${\cal N}_{Kijk}$ is defined as 
\beq
{\cal N}_{Kijk} = h_x h_y h_z(h_x u_i+h_y v_j)(h_x u_i +h_z w_k)(h_y v_j+h_z w_k) {\cal R}_{Kijk}^2
\eeqn{5.3}
where ${\cal R}_{Kijk} = {\cal R}_K(h_x u_i,h_y v_j,h_z w_k)$. 

The functions $F^K_{ijk}(x,y,z)$ satisfy the Lagrange property with respect to the three-dimensional mesh 
$(h_x u_p,h_y v_q,h_z w_r)$, i.e. they vanish at all mesh points but one, 
\beq
F^K_{ijk}(h_x u_p,h_y v_q,h_z w_r)=({\cal N}_{Kijk} {\cal R}_{Kijk}^{-2} 
\lambda_i^{N_x} \lambda_j^{N_y} \lambda_k^{N_z})^{-1/2} \delta_{ip}\delta_{jq}\delta_{kr}.
\eeqn{5.4}
The coefficients $\lambda_i^{N_x}$, $\lambda_j^{N_y}$, $\lambda_k^{N_z}$ are the Christoffel numbers 
which appear as weights in the Gauss-Laguerre quadrature approximation
\beq
\int_{0}^{\infty} \int_{0}^{\infty} \int_{0}^{\infty} G(u,v,w) du dv dw 
\approx \sum_{i=1}^{N_x} \sum_{j=1}^{N_y} \sum_{k=1}^{N_z} 
\lambda_i^{N_x} \lambda_j^{N_y} \lambda_k^{N_z} G(u_i,v_j,w_k).
\eeqn{5.7}
At the Gauss approximation scaled with $h_x$, $h_y$, $h_z$, 
the three-dimensional Lagrange functions \eref{5.1} are orthonormal 
with respect to the perimetric volume element because of the Lagrange property \eref{5.4}. 

The $\Phi_K^{(L^{\pi})\sigma}(x,y,z)$ functions of equation \eref{4.1} are expanded in the Lagrange basis as
\beq
\Phi_K^{(L^{\pi})\sigma}(x,y,z) &=& \sum_{i=1}^{N} \sum_{j=1}^{i-\delta_K} \sum_{k=1}^{N_z} 
C_{Kijk}^{(L^{\pi})\sigma} \left[ 2 (1+\delta_{ij})\right]^{-1/2}\nonumber\\
&& \times \left[ F^K_{ijk}(x,y,z) + \sigma \pi (-1)^K F^K_{jik}(x,y,z) \right]
\eeqn{5.5}
where we use the same number $N$ of mesh points and the same scale factor $h$ for the two perimetric 
coordinates $x$ and $y$ in order to take advantage of the Lagrange conditions \eref{5.4} 
when the two coordinates are exchanged. 
Because of the symmetrization the sum over $j$ is limited by the value $i-\delta_K$, 
where $\delta_K$ is equal to 0 when $(-1)^K=\sigma \pi$ and to 1 when $(-1)^K=-\sigma \pi$. 

The three-body Hamiltonian in perimetric coordinates for each good quantum number $L$ 
and its discretization on a Lagrange mesh are given in \cite{HB01}. 
The singularities of the three Coulomb terms are automatically regularized in the matrix elements 
by the volume element so that the Lagrange-mesh method is not affected by those singularities. 
The potential matrix is diagonal and its elements are the values of the potential at the mesh points. 
The calculation would be as easy with other form factors for the potentials. 
The resulting matrix is rather sparse. 
The remaining problem is to calculate the lowest eigenvalues and the corresponding eigenvectors 
of a large sparse matrix. 

For given $L^\pi$, the eigenvalues in increasing order are labeled by a quantum number $v \ge 0$ 
related to the vibrational excitation in the Born-Oppenheimer picture. 
The corresponding eigenvectors provide the coefficients appearing in expansion \eref{5.5}. 

Let us consider initial and final components \eref{5.5} 
with respective coefficients $C_{K_iijk}^{i(L_i^{\pi_i})\sigma_i}$ and 
$C_{K_fijk}^{f(L_f^{\pi_f})\sigma_f}$. 
Because of the Lagrange property \eref{5.4}, 
the matrix elements \eref{4.6} are simply obtained with the Gauss quadrature \eref{5.7} as 
\beq
A_{K_i K_f;\kappa}^{L_i L_f} 
\approx \sum_{i=1}^N \sum_{j=1}^{i-\delta_K} \sum_{k=1}^{N_z} 
C_{K_iijk}^{i(L_i^{\pi_i})\sigma_i} C_{K_fijk}^{f(L_f^{\pi_f})\sigma_f} 
\mathcal{A_\kappa}(h u_i,h v_j,h_z w_k)
\eeqn{5.6}
where $\delta_K = \max(\delta_{K_i},\delta_{K_f})$.

\section{Energies and E2 transition probabilities}
\label{sec:res}
\subsection{H$_2^+$ bound and quasibound energies}
The energies of the $v = 0$ lowest vibrational bound states for $L=0$ to 35 have been 
calculated with the present method in \cite{HB03}. 
Here we extend those calculations to quasibound states up to $L = 40$ 
and to the first three excited vibrational states. 
The main reason making this extension possible is a better technique for searching 
the eigenvalues of a large sparse matrix \cite{BN07} and faster personal computers. 

Since the main aim is to calculate transition matrix elements 
involving two different wave functions, 
it is convenient to use a single three-dimensional mesh for all states. 
An excellent accuracy is obtained when the parameters of the calculation are chosen 
as $N = N_x = N_y = 40$, $N_z = 20$ and $h = h_x = h_y = 0.14$, $h_z = 0.4$. 
For a given $K$ value, the total number of basis states is then 16400 or 15600 
depending on the parity of $K$. 
The size of the matrix is larger by about a factor ($K_{\rm max} + 1)$ 
when $K$ is limited by $K \le K_{\rm max} \le L$. 
For $K>2$, like in \cite{HB03}, calculations are performed with $K_{\rm max} = 2$. 
They correspond to a size of 48400. 
In order to make comparisons with more accurate literature results, 
we use for the proton mass the benchmark value $m_p = 1836.152701$ a.u. 
The dissociation threshold $E_d$ is then at $-0.499\,727\,839\,716$ a.u.\ or Hartrees. 
The obtained energies are presented as the first line for each $L$ value in Table \ref{tab:1}. 
The accuracy is estimated from the stability of the digits with respect to calculations 
with $N \pm 2$ mesh points. 
The error is expected to be at most of a few units on the last displayed digit. 
Literature results sometimes truncated at the 15th digit are displayed in each second line. 
Except in the low-$L$ or low $v$ regions where other references are mentioned, 
the literature results are the 11-digit energies obtained by Moss \cite{Mo93}. 

\begin{figure}[hbt]
\setlength{\unitlength}{1mm}
\begin{picture}(140,70) (-20,10) 
\put(0,0){\mbox{\scalebox{1.5}{\includegraphics{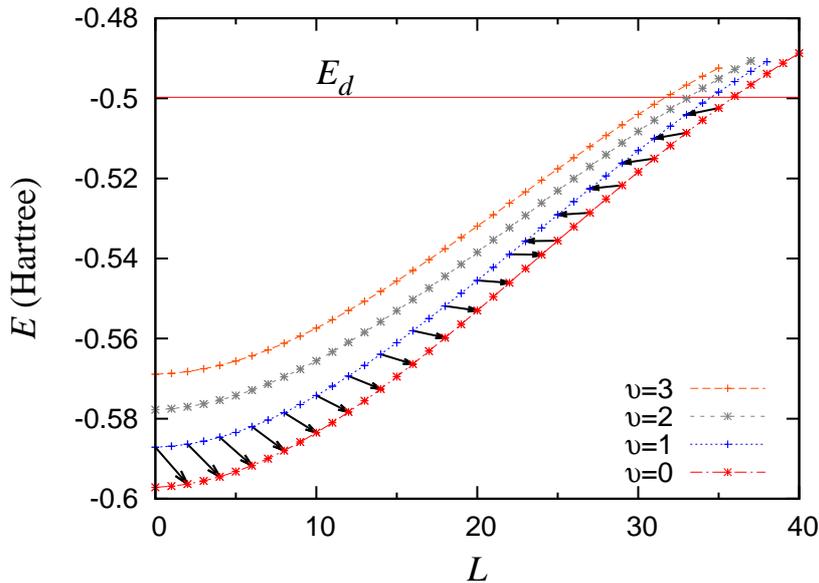}}}}
\end{picture} \\
\caption{Four lowest $\Sigma_g$ rotational bands of the H$_2^+$ molecular ion 
and dissociation energy $E_d$. 
Arrows show how the direction of $L \rightarrow L+2$ transitions between the two lowest 
bands changes along the band.}
\label{fig:1}
\end{figure}

\begin{center}
\begin{longtable}{rllll}
\caption{Energies of the four lowest vibrational bound or quasibound states 
in the $\Sigma_g$ rotational band of the H$_2^+$ molecular ion. 
Quasibound states are separated from bound states by a horizontal bar. 
For each $L$ value, the Lagrange-mesh results obtained with $N_x = N_y = 40$, $N_z = 20$ and 
$h_x = h_y = 0.14$, $h_z = 0.4$ are presented in the first line. 
The second line displays the results of Moss \cite{Mo93} 
except when other references are indicated ($^a$: \cite{HNN09}, $^b$: \cite{LWZ07}, 
$^c$: \cite{HBG00}, $^d$: \cite{YZ04}). 
For $L=0$, Lagrange-mesh results obtained with $N = 55$, $N_z = 25$ 
are given in the third line. 
The proton mass is taken as $m_p=1836.152701 m_e$.}
\label{tab:1}\\
\hline %Header for the first page
$L$& $v=0$          & $v=1$           &  $v=2$          &$v=3$\\
\hline
\endfirsthead
\multicolumn{5}{c}{{\tablename} \thetable{} -- Continuation}\\ %header for the remaining
\hline
$L$& $v=0$          & $v=1$           &  $v=2$          &$v=3$\\
\hline
\endhead
\hline
\multicolumn{5}{l}{{Continued on Next Page\ldots}}\\%Footer for all pages except the last page
\endfoot
\hline                                       %Footer for the last page
\endlastfoot
{0}
&-0.597\,139\,063\,123\,3 &-0.587\,155\,679\,207&-0.577\,751\,904\,49&-0.568\,908\,497\,8 \\
&-0.597\,139\,063\,123\,405$^{a,b}$
&-0.587\,155\,679\,212\,747$^b$
&-0.577\,751\,904\,595\,47$^{c}$
&-0.568\,908\,498\,966\,77$^{c}$\\
&-0.597\,139\,063\,123\,41
&-0.587\,155\,679\,212\,76&-0.577\,751\,904\,595\,47&-0.568\,908\,498\,966\,75\\
{1}     
&-0.596\,873\,738\,832\,6&-0.586\,904\,321\,034&-0.577\,514\,034\,14&-0.568\,683\,707\,4 \\
&-0.596\,873\,738\,832\,765$^{a,b}$
&-0.586\,904\,321\,04$^{c}$
&-0.577\,514\,034\,24$^{c}$
&-0.568\,683\,708\,50$^{c}$\\
{2}     
&-0.596\,345\,205\,545\,3&-0.586\,403\,631\,650&-0.577\,040\,237\,25&-0.568\,235\,992\,1 \\
&-0.596\,345\,205\,545\,46$^{d}$ &-0.586\,403\,631\,64 &-0.577\,040\,237\,32&-0.568\,235\,993\,18\\
{3}     
&-0.595\,557\,639\,048\,1&-0.585\,657\,612\,010&-0.576\,334\,350\,31&-0.567\,569\,033\,9 \\
&-0.595\,557\,639\,048\,23$^{d}$ &-0.585\,657\,612\,02 &-0.576\,334\,350\,40&-0.567\,569\,035\,05\\
{4}     
&-0.594\,517\,169\,322\,3&-0.584\,672\,134\,376&-0.575\,402\,003\,40&-0.566\,688\,235\,7 \\
&-0.594\,517\,169\,322\,41$^{d}$ &-0.584\,672\,134\,39 &-0.575\,402\,003\,51&-0.566\,688\,236\,89\\
{5}       
&-0.593\,231\,728\,998\,2&-0.583\,454\,796\,107&-0.574\,250\,479\,24&-0.565\,600\,585\,3 \\
&-0.593\,231\,728\,998\,34$^{d}$ &-0.583\,454\,796\,10 &-0.574\,250\,479\,35&-0.565\,600\,586\,54\\
{6}       
&-0.591\,710\,865\,102\,0&-0.582\,014\,738\,323&-0.572\,888\,538\,69&-0.564\,314\,488\,1 \\
&-0.591\,710\,865\,102\,11$^{d}$ &-0.582\,014\,738\,32 &-0.572\,888\,538\,80&-0.564\,314\,489\,33\\
{7}       
&-0.589\,965\,524\,077\,8&-0.580\,362\,439\,246&-0.571\,326\,222\,24&-0.562\,839\,576\,1 \\
&-0.589\,965\,524\,077\,97$^{d}$ &-0.580\,362\,439\,24 &-0.571\,326\,222\,33&-0.562\,839\,577\,36\\
{8}     
&-0.588\,007\,820\,606\,9&-0.578\,509\,492\,482&-0.569\,574\,637\,54&-0.561\,186\,505\,1 \\
&-0.588\,007\,820\,606\,99$^{d}$ &-0.578\,509\,492\,46 &-0.569\,574\,637\,56&-0.561\,186\,506\,35\\
{9}       
&-0.585\,850\,800\,433\,7&-0.576\,468\,380\,169&-0.567\,645\,742\,58&-0.559\,366\,747\,1 \\
&-0.585\,850\,800\,433\,87$^{d}$ &-0.576\,468\,380\,15 &-0.567\,645\,742\,64&-0.559\,366\,748\,32\\
{10}       
&-0.583\,508\,206\,414\,4&-0.574\,252\,249\,872&-0.565\,552\,133\,03&-0.557\,392\,388\,2 \\
&-0.583\,508\,206\,414\,57$^{d}$ &-0.574\,252\,249\,85 &-0.565\,552\,133\,10&-0.557\,392\,389\,42\\
{11}       
&-0.580\,994\,255\,517\,9&-0.571\,874\,702\,601&-0.563\,306\,840\,95&-0.555\,275\,937\,7 \\
&-0.580\,994\,255\,518\,01$^{d}$ &-0.571\,874\,702\,58 &-0.563\,306\,841\,02&-0.555\,275\,938\,84\\
{12}     
&-0.578\,323\,432\,777\,8&-0.569\,349\,597\,646&-0.560\,923\,150\,05&-0.553\,030\,153\,2 \\
&-0.578\,323\,432\,777\,97$^{d}$ &-0.569\,349\,597\,61 &-0.560\,923\,150\,12&-0.553\,030\,154\,22\\
{13}       
&-0.575\,510\,306\,398\,5&-0.566\,690\,878\,119&-0.558\,414\,431\,31&-0.550\,667\,886\,2 \\
&-0.575\,510\,306\,37    &-0.566\,690\,878\,11 &-0.558\,414\,431\,38&-0.550\,667\,887\,26\\
{14}       
&-0.572\,569\,366\,530\,4&-0.563\,912\,419\,513&-0.555\,794\,001\,00&-0.548\,201\,949\,8 \\
&-0.572\,569\,366\,51    &-0.563\,912\,419\,49 &-0.555\,794\,001\,09&-0.548\,201\,950\,86\\
{15}       
&-0.569\,514\,888\,762\,2&-0.561\,027\,902\,144&-0.553\,075\,001\,79&-0.545\,645\,007\,9 \\
&-0.569\,514\,888\,74    &-0.561\,027\,902\,14 &-0.553\,075\,001\,87&-0.545\,645\,008\,95\\
{16}     
&-0.566\,360\,822\,174\,0&-0.558\,050\,707\,254&-0.550\,270\,306\,85&-0.543\,009\,487\,9 \\
&-0.566\,360\,822\,15    &-0.558\,050\,707\,24 &-0.550\,270\,306\,91&-0.543\,009\,488\,90\\
{17}       
&-0.563\,120\,700\,897\,8&-0.554\,993\,835\,680&-0.547\,392\,445\,68&-0.540\,307\,513\,7 \\
&-0.563\,120\,700\,88    &-0.554\,993\,835\,67 &-0.547\,392\,445\,73&-0.540\,307\,514\,61\\
{18}       
&-0.559\,807\,577\,501\,8&-0.551\,869\,847\,486&-0.544\,453\,550\,24&-0.537\,550\,859\,0 \\
&-0.559\,807\,577\,45    &-0.551\,869\,847\,45 &-0.544\,453\,550\,29&-0.537\,550\,859\,83\\
{19}       
&-0.556\,433\,976\,142\,5&-0.548\,690\,820\,585&-0.541\,465\,319\,58&-0.534\,750\,919\,3 \\
&-0.556\,433\,976\,10    &-0.548\,690\,820\,57 &-0.541\,465\,319\,63&-0.534\,750\,920\,04\\
{20}     
&-0.553\,011\,863\,258\,8&-0.545\,468\,326\,311&-0.538\,439\,001\,16&-0.531\,918\,700\,1 \\
&-0.553\,011\,863\,22    &-0.545\,468\,326\,30 &-0.538\,439\,001\,20&-0.531\,918\,700\,81\\
{21}       
&-0.549\,552\,633\,572\,9&-0.542\,213\,419\,906&-0.535\,385\,387\,00&-0.529\,064\,821\,1 \\
&-0.549\,552\,633\,53    &-0.542\,213\,419\,86 &-0.535\,385\,387\,04&-0.529\,064\,821\,72\\
{22}       
&-0.546\,067\,109\,285\,0&-0.538\,936\,644\,076&-0.532\,314\,823\,38&-0.526\,199\,534\,8 \\
&-0.546\,067\,109\,22    &-0.538\,936\,644\,05 &-0.532\,314\,823\,41&-0.526\,199\,535\,29\\
{23}       
&-0.542\,565\,550\,550\,5&-0.535\,648\,044\,025&-0.529\,237\,232\,79&-0.523\,332\,759\,6 \\
&-0.542\,565\,550\,49    &-0.535\,648\,044\,01 &-0.529\,237\,232\,81&-0.523\,332\,759\,99\\
{24}     
&-0.539\,057\,675\,600\,7&-0.532\,357\,192\,725&-0.526\,162\,147\,73&-0.520\,474\,127\,9 \\
&-0.539\,057\,675\,58    &-0.532\,357\,192\,68 &-0.526\,162\,147\,74&-0.520\,474\,128\,23\\
{25}       
&-0.535\,552\,689\,189\,4&-0.529\,073\,225\,592&-0.523\,098\,756\,19&-0.517\,633\,051\,1 \\
&-0.535\,552\,689\,16    &-0.529\,073\,225\,56 &-0.523\,098\,756\,18&-0.517\,633\,051\,39\\
{26}       
&-0.532\,059\,318\,403\,4&-0.525\,804\,884\,277&-0.520\,055\,959\,88&-0.514\,818\,804\,7 \\
&-0.532\,059\,318\,34    &-0.525\,804\,884\,21 &-0.520\,055\,959\,86&-0.514\,818\,804\,90\\
{27}       
&-0.528\,585\,855\,292\,2&-0.522\,560\,569\,912&-0.517\,042\,447\,52&-0.512\,040\,640\,1 \\
&-0.528\,585\,855\,25    &-0.522\,560\,569\,84 &-0.517\,042\,447\,51&-0.512\,040\,640\,25\\
{28}     
&-0.525\,140\,206\,242\,8&-0.519\,348\,407\,044&-0.514\,066\,787\,02&-0.509\,307\,935\,1 \\
&-0.525\,140\,206\,16    &-0.519\,348\,407\,01 &-0.514\,066\,787\,00&-0.509\,307\,935\,21\\
{29}       
&-0.521\,729\,948\,617\,0&-0.516\,176\,320\,684&-0.511\,137\,543\,92&-0.506\,630\,401\,9 \\
&-0.521\,729\,948\,55    &-0.516\,176\,320\,61 &-0.511\,137\,543\,89&-0.506\,630\,401\,96\\
{30}       
&-0.518\,362\,395\,936\,9&-0.513\,052\,130\,786&-0.508\,263\,438\,21&-0.504\,018\,389\,9 \\
&-0.518\,362\,395\,87    &-0.513\,052\,130\,73 &-0.508\,263\,438\,18&-0.504\,018\,389\,91\\
{31}       
&-0.515\,044\,673\,983\,8&-0.509\,983\,671\,425&-0.505\,453\,561\,05&-0.501\,483\,354\,1 \\
&-0.515\,044\,673\,92    &-0.509\,983\,671\,35 &-0.505\,453\,560\,98&-0.501\,483\,354\,05\\\cline{5-5}
{32}     
&-0.511\,783\,811\,807\,3&-0.506\,978\,947\,094&-0.502\,717\,691\,44&-0.499\,038\,641\,7 \\
&-0.511\,783\,811\,71    &-0.506\,978\,947\,03 &-0.502\,717\,691\,38&-0.499\,038\,641\,65\\
{33}       
&-0.508\,586\,854\,260\,8&-0.504\,046\,347\,959&-0.500\,066\,791\,73&-0.496\,700\,970\,5 \\
&-0.508\,586\,854\,19    &-0.504\,046\,347\,86 &-0.500\,066\,791\,69&-0.496\,700\,970\,47\\\cline{4-4}
{34}       
&-0.505\,461\,007\,173\,3&-0.501\,194\,964\,711&-0.497\,513\,855\,93&-0.494\,493\,731\,1 \\
&-0.505\,461\,007\,08    &-0.501\,194\,964\,64 &-0.497\,513\,855\,86&-0.494\,493\,731\,12\\\cline{3-3}
{35}       
&-0.502\,413\,834\,509\,9&-0.498\,435\,084\,588&-0.495\,075\,553\,29&-0.492\,457         \\
&-0.502\,413\,834\,44    &-0.498\,435\,084\,52 &-0.495\,075\,553\,22&-0.492\,457\,8      \\\cline{2-2}
{36}     
&-0.499\,453\,543\,217\,6&-0.495\,779\,051\,406&-0.492\,776\,112    & \\
&-0.499\,453\,543\,14    &-0.495\,779\,051\,34 &-0.492\,776\,110\,97& \\
{37}       
&-0.496\,589\,427\,076\,0&-0.493\,242\,971\,555&-0.490\,66          & \\
&-0.496\,589\,426\,96    &-0.493\,242\,971\,49 &-0.490\,660\,7      & \\
{38}       
&-0.493\,832\,629\,114\,0&-0.490\,850\,95      & & \\
&-0.493\,832\,629\,02    &-0.490\,850\,96      & & \\
{39}       
&-0.491\,197\,646\,03    & & & \\
&-0.491\,197\,645\,93    & & & \\
{40}
&-0.488\,706\,1          & & & \\
&-0.488\,706\,110        & & & \\
\end{longtable}
\end{center}             
Let us start the discussion with the $(L^\pi,v) = (0^+,0)$ ground state. 
The energy of the $0^+$ ground state has been improved with respect to Moss' result 
in a number of papers \cite{Ko00,HBG00,BF02,YZL03,CD04,LWZ07,HNN09} 
to reach an accuracy around or beyond 30 digits in \cite{LWZ07,HNN09}. 
Our accuracy is about $10^{-13}$. 
For the $(0^+,1)$ first vibrational excited state, the accuracy is better than $10^{-11}$ \cite{LWZ07}. 
For the $(0^+,2)$ and $(0^+,3)$ states, the accuracies are about $10^{-10}$ and $10^{-9}$, 
respectively \cite{HBG00}. 
There is thus about a one-digit loss for each vibrational excitation. 
When the numbers of mesh points are increased to $N = 55$ and $N_z = 25$, 
the accuracies of all these states reach about $10^{-14}$ (see the third line of the $L = 0$ energies). 
The energy of the lowest $L = 1$ state is known with about 25 digits \cite{LWZ07,HNN09}. 
Our result has an accuracy better than $2 \times 10^{-13}$. 
Comparisons with references \cite{Mo93,HBG00} indicate that the accuracies 
of the $L = 1$ vibrational excited states behave like for $L = 0$. 
Results with an accuracy close to 18 digits are available for $L = 2$-12 and $v = 0$ \cite{YZ04}. 
They show that our error for the lowest vibrational energy remains smaller than $2 \times 10^{-13}$ 
for all these states. 
When comparing the rest of our results with those of Moss \cite{Mo93}, 
one observes that both works agree very well. 
We are a little more accurate for $v=0$ and a little less accurate for $v=2$ and 3. 
But in addition, the Lagrange-mesh method provides easy-to-use wave functions. 
The obtained spectrum is depicted in Fig.~\ref{fig:1}. 
\begin{table}[hbt]
\centering{
\caption{Convergence of the energies and transition probabilities 
as a function of the numbers $N$ and $N_z$ of mesh points. 
Two cases are shown: $(4^+,0) \rightarrow (2^+,0)$ where $L_f = L_i + 2$ (upper set) 
and $(30^+,2) \rightarrow (32^+,0)$ where $L_f = L_i - 2$ (lower set). 
The scale factors are $h = 0.14$ and $h_z = 0.4$.} 
\label{tab:2} 
\resizebox{16cm}{!}{
\begin{tabular}{lllllll}
\hline
$N$&$N_z$&$E_i (4^+,0)$&$E_f (2^+,0)$&$W_0$ ($10^{-10}$s$^{-1}$)&$W_1$ ($10^{-10}$s$^{-1}$)&$W$ ($10^{-10}$s$^{-1}$)\\
\hline
20&20&-0.594\,516\,6           &-0.596\,344\,6           &9.213\,5       &9.209\,8       &9.209\,8       \\
25&20&-0.594\,517\,156\,       &-0.596\,345\,193\,       &9.212\,254\,   &9.208\,480\,   &9.208\,477\,   \\
30&20&-0.594\,517\,169\,03     &-0.596\,345\,205\,26     &9.212\,220\,3  &9.208\,445\,8  &9.208\,442\,3  \\
35&20&-0.594\,517\,169\,315\,  &-0.596\,345\,205\,539\,  &9.212\,219\,410&9.208\,444\,904&9.208\,441\,446\\
35&30&-0.594\,517\,169\,315\,  &-0.596\,345\,205\,539\,  &9.212\,219\,411&9.208\,444\,904&9.208\,441\,437\\
40&20&-0.594\,517\,169\,322\,29&-0.596\,345\,205\,545\,33&9.212\,219\,383&9.208\,444\,877&9.208\,441\,409\\
40&30&-0.594\,517\,169\,322\,30&-0.596\,345\,205\,545\,33&9.212\,219\,384&9.208\,444\,878&9.208\,441\,410\\
\cite{YZ04}&  &-0.594\,517\,169\,322\,41&-0.596\,345\,205\,545\,46&&&\\
\hline
$N$&$N_z$&$E_i (30^+,2)$&$E_f (32^+,0)$&$W_0$ ($10^{-10}$s$^{-1}$)&$W_1$ ($10^{-10}$s$^{-1}$)&$W$ ($10^{-10}$s$^{-1}$)\\
\hline
20&20&-0.508\,07               &-0.511\,781\,9           &2.72           &2.72           &2.72           \\
25&20&-0.508\,260\,8           &-0.511\,783\,798\,       &2.394\,0       &2.393\,5       &2.393\,5       \\
30&20&-0.508\,263\,395\,       &-0.511\,783\,811\,69     &2.391\,953\,   &2.391\,480\,   &2.391\,488\,   \\
35&20&-0.508\,263\,437\,6      &-0.511\,783\,811\,805\,9 &2.391\,930\,599&2.391\,457\,162&2.391\,465\,808\\
35&30&-0.508\,263\,437\,6      &-0.511\,783\,811\,805\,9 &2.391\,930\,596&2.391\,457\,159&2.391\,465\,806\\
40&20&-0.508\,263\,438\,21     &-0.511\,783\,811\,807\,25&2.391\,930\,370&2.391\,456\,933&2.391\,465\,579\\
40&30&-0.508\,263\,438\,21     &-0.511\,783\,811\,807\,25&2.391\,930\,368&2.391\,456\,931&2.391\,465\,577\\
\cite{Mo93}&  &-0.508\,263\,438\,18&-0.511\,783\,811\,71&&&\\
\hline
\end{tabular}
}}
\end{table}
Typical convergence tests are displayed in Table \ref{tab:2} for two quite different 
sets of initial and final states. 
For the values $h = 0.14$ and $h_z = 0.4$ close to those suggested in \cite{HB03}, 
Table \ref{tab:2} displays initial and final energies 
and transition probabilities for various choices of $N$ and $N_z$. 
The transition probability $W_0$ is obtained by restricting \eref{4.8} to $\kappa = 0$ 
while $W_1$ corresponds to $\kappa \le 1$. 
First one observes that the $\kappa = 1$ contributions have an importance smaller than 0.05 \% 
and that the $\kappa = 2$ contributions are smaller than 0.001 \%. 
Second one sees that the convergence of the transition probabilities is slower than 
the convergence of the energies, as expected from the variational principle. 
A good convergence with respect to $N_z$ is already obtained for $N_z = 20$. 
The convergence with respect to $N$ is slower. 
Increasing $N$ is more expensive than increasing $N_z$ 
since the size of the basis increases with $N^2$. 
Since the convergence is exponential, one can estimate that 
the relative accuracy on $W$ is about $10^{-9}$ for $(4^+,0)\rightarrow(2^+,0)$ 
and still better than $10^{-8}$ for $(30^+,2)\rightarrow(32^+,0)$. 
This means that the wave functions are quite accurate. 
Further similar tests have been performed for other transitions. 

The convergence of the transition probabilities with respect to $K_{\rm max}$ can be studied 
by comparison with results from wave functions truncated at $K_{\rm max} = 0$ 
and $K_{\rm max}= 1$. 
The relative error when $K_{\rm max} = 0$ is smaller than 0.3 \% for all considered transitions 
while the error for $K_{\rm max} = 1$ is smaller than $10^{-5}$. 
This is rather similar to truncations with respect to $\kappa$. 
By extrapolation, we estimate that the relative error on the present transition probabilities 
obtained with $K_{\rm max}= 2$ should be smaller than $10^{-7}$. 

The probabilities per second for transitions within a same rotational band, $L_f = L_i - 2$ 
and $v_f = v_i \le 3$, are presented in Table \ref{tab:3}. 
They include some transition probabilities involving quasibound states. 
In accord with the previous discussion, we limit the number of significant figures to six. 
The probabilities increase slowly with $L$ with a maximum around $L_i = 32$ for $v_i = 0$, 
$L_i = 30$ for $v_i = 1$, $L_i = 28$ for $v_i = 2$, and $L_i = 27$ for $v_i = 3$, 
not far from the end of the rotational bands. 
This is due to a maximum of the energy differences around $L_i = 25$. 
The maximum of the transition probabilities is shifted toward higher $L_i$ values 
by a steady increase of the reduced matrix elements.
The transition probabilities behave similarly for the displayed $v$ values 
with a slight decrease when $v$ increases. 
\begin{center}
\begin{longtable}{rllll}
\caption{Quadrupole transition probabilities per second $W$ for transitions 
between states of a same rotational band ($v_f = v_i$, $L_f = L_i - 2$). 
Results are given with five digits followed by the power of 10. } 
\label{tab:3} \\
\\[-4.9ex]
\hline
$L_i$&$v_i=0$&$v_i=1$&$v_i=2$&$v_i=3$\\
\hline
\endfirsthead
\multicolumn{4}{c}{{\tablename} \thetable{} -- Continuation}\\
\hline
$L_i$&$v_i=0$&$v_i=1$&$v_i=2$&$v_i=3$\\
\hline
\endhead
\hline
\multicolumn{4}{l}{{Continued on Next Page\ldots}}\\
\endfoot
\hline
\endlastfoot
%The data...
 2&9.731\,37-12&9.680\,53-12&9.441\,58-12&9.038\,44-12\\
 3&1.581\,33-10&1.570\,47-10&1.529\,34-10&1.461\,88-10\\
 4&9.208\,44-10&9.122\,85-10&8.863\,66-10&8.454\,10-10\\
 5&3.316\,04-09&3.274\,74-09&3.172\,19-09&3.016\,92-09\\
 6&9.003\,66-09&8.856\,98-09&8.548\,31-09&8.101\,26-09\\
 7&2.027\,18-08&1.985\,14-08&1.907\,78-08&1.800\,55-08\\
 8&3.991\,47-08&3.888\,76-08&3.719\,19-08&3.493\,66-08\\
 9&7.102\,43-08&6.880\,86-08&6.545\,68-08&6.116\,68-08\\
10&1.167\,28-07&1.124\,01-07&1.063\,06-07&9.877\,20-08\\
11&1.799\,09-07&1.721\,22-07&1.617\,76-07&1.493\,87-07\\
12&2.629\,44-07&2.498\,55-07&2.332\,90-07&2.140\,06-07\\
13&3.674\,92-07&3.467\,23-07&3.214\,91-07&2.928\,58-07\\
14&4.943\,34-07&4.629\,69-07&4.261\,64-07&3.853\,47-07\\
15&6.432\,99-07&5.979\,13-07&5.462\,22-07&4.900\,73-07\\
16&8.132\,46-07&7.499\,75-07&6.797\,55-07&6.049\,01-07\\
17&1.002\,12-06&9.167\,53-07&8.241\,37-07&7.270\,87-07\\
18&1.207\,04-06&1.095\,14-06&9.761\,56-07&8.534\,13-07\\
19&1.424\,45-06&1.281\,49-06&1.132\,17-06&9.803\,47-07\\
20&1.650\,27-06&1.471\,75-06&1.288\,26-06&1.104\,19-06\\
21&1.880\,04-06&1.661\,62-06&1.440\,38-06&1.221\,22-06\\
22&2.109\,12-06&1.846\,74-06&1.584\,53-06&1.327\,83-06\\
23&2.332\,81-06&2.022\,80-06&1.716\,79-06&1.420\,58-06\\
24&2.546\,48-06&2.185\,62-06&1.833\,51-06&1.496\,35-06\\
25&2.745\,72-06&2.331\,32-06&1.931\,34-06&1.552\,32-06\\
26&2.926\,40-06&2.456\,32-06&2.007\,25-06&1.586\,08-06\\
27&3.084\,74-06&2.557\,42-06&2.058\,62-06&1.595\,57-06\\
28&3.217\,37-06&2.631\,82-06&2.083\,22-06&1.579\,13-06\\
29&3.321\,35-06&2.677\,12-06&2.079\,17-06&1.535\,44-06\\
30&3.394\,20-06&2.691\,32-06&2.044\,95-06&1.463\,47-06\\
31&3.433\,83-06&2.672\,76-06&1.979\,30-06&1.362\,40-06\\
32&3.438\,58-06&2.620\,08-06&1.881\,15-06&1.231\,41-06\\
33&3.407\,11-06&2.532\,10-06&1.749\,39-06&1.069\,27-06\\
34&3.338\,37-06&2.407\,67-06&1.582\,54-06&8.732\,11-07\\
35&3.231\,44-06&2.245\,36-06&1.377\,99-06&6.32\ \ \ \ \ -07\\
36&3.085\,37-06&2.042\,91-06&1.129\,4\,\ -06&   \\
37&2.898\,82-06&1.795\,86-06&8.1\,\ \ \ \ \ \ -07&   \\
38&2.669\,34-06&1.492\,9\,\ -06&   &   \\
39&2.391\,63-06&   &   &   \\
40&2.051\,\ \ \ -06&   &   &   \\
\end{longtable}
%}
\end{center}

The probabilities per second for other transitions are displayed in Table \ref{tab:4}. 
The columns correspond to transitions between different vibrational states. 
For each $L_i$ value, the successive lines correspond to increasing values of $L_f$, 
i.e., to $L_f = L_i - 2$ for $L_i > 1$, $L_f = L_i$ for $L_i > 0$, and $L_f = L_i + 2$, respectively. 
For $L_i \le 20$, the obtained probabilities agree with those of Posen \etal \cite{PDP83} 
and improve them significantly. 
More than 95 \% of the results of Posen \etal exactly correspond to our result 
rounded at two significant figures. 
In most other cases, the rounding is of 6 units on the third digit rather than at most 5 
for a normal rounding. 
An example of the few 'worst' cases is the $(19^-,3) \rightarrow (17^-,2)$ transition probability 
where our result in Table \ref{tab:4} is $5.847\,00 \times 10^{-10}$ while the result 
of Posen \etal is $5.7 \times 10^{-10}$. 

The strongest transition from each state occurs in general towards the nearest vibrational state 
($v_f = v_i - 1$) for $L_f = L_i - 2$. 
Exceptions can be found between $L_i = 14$ and $L_i = 22$. 
For $v_f = v_i - 1$, in the vicinity of $L_i = 23$ and beyond, 
the  $(L_i,v_i) \rightarrow (L_i + 2,v_i-1)$ transitions 
are replaced by $(L_i + 2,v_i-1) \rightarrow (L_i,v_i)$ transitions 
because the sign of the energy difference changes 
(see the arrows in Fig.~\ref{fig:1} for the $1 \rightarrow 0$ transitions). 
These numbers are indicated in italics in Table \ref{tab:4}. 
For example, the first number in the last line for $L_i = 23$ 
corresponds to the $(25,0) \rightarrow (23,1)$ transition. 
Hence the transition probabilities strongly vary in this region. 
\begin{center}
\begin{longtable}{lllllll}
\caption{Quadrupole transition probabilities per second $W$ for transitions 
between different vibrational quantum numbers $(v_i \ne v_f)$. 
The three successive lines correspond to increasing $L_f$ values, i.e.\ 
$L_f = L_i - 2$, $L_f = L_i$ and $L_f = L_i + 2$, respectively, for $L_i \ge 2$. 
Italicized numbers for $(1 \rightarrow 0)$, $(2 \rightarrow 1)$ and $(3 \rightarrow 2)$ 
mean that the initial and final states are exchanged 
(the first one is preceded in each case by a horizontal bar).} 
\label{tab:4}\\
\\[-4.9ex]
\hline
$L_i$&$(1\rightarrow\,0)$&$(2\rightarrow\,0)$&$(2\rightarrow\,1)$
     &$(3\rightarrow\,0)$&$(3\rightarrow\,1)$&$(3\rightarrow\,2)$\\
\hline
\endfirsthead
\multicolumn{7}{c}{{\tablename} \thetable{} -- Continuation}\\
\hline
$L_i$&$(1\rightarrow\,0)$&$(2\rightarrow\,0)$&$(2\rightarrow\,1)$
     &$(3\rightarrow\,0)$&$(3\rightarrow\,1)$&$(3\rightarrow\,2)$\\
\hline
\endhead
\hline
\multicolumn{7}{l}{{Continued on Next Page\ldots}}\\
\endfoot
\hline
\endlastfoot
%The data...
 0&5.215\,07-07&6.777\,83-08&8.571\,80-07&6.277\,98-09&1.780\,77-07&1.045\,81-06\\
 1&2.649\,11-07&4.310\,45-08&4.326\,10-07&5.534\,79-09&1.103\,90-07&5.244\,04-07\\
  &2.586\,05-07&2.801\,60-08&4.263\,10-07&1.845\,73-09&7.518\,42-08&5.215\,85-07\\
 2&1.602\,07-07&3.128\,77-08&2.593\,97-07&4.978\,71-09&7.845\,99-08&3.117\,14-07\\
  &1.878\,56-07&3.070\,72-08&3.066\,23-07&3.960\,55-09&7.859\,99-08&3.714\,70-07\\
  &1.788\,44-07&1.552\,08-08&2.953\,94-07&5.794\,07-10&4.279\,53-08&3.620\,35-07\\
 3&2.266\,89-07&4.932\,77-08&3.644\,12-07&8.766\,41-09&1.221\,20-07&4.346\,60-07\\
  &1.734\,34-07&2.854\,37-08&2.828\,71-07&3.705\,83-09&7.300\,64-08&3.423\,98-07\\
  &1.303\,97-07&8.543\,49-09&2.155\,75-07&9.058\,65-11&2.444\,93-08&2.643\,79-07\\
 4&2.707\,32-07&6.511\,44-08&4.315\,88-07&1.270\,26-08&1.592\,35-07&5.102\,93-07\\
  &1.664\,90-07&2.764\,82-08&2.712\,73-07&3.620\,52-09&7.064\,46-08&3.279\,79-07\\
  &9.561\,75-08&4.302\,52-09&1.580\,64-07&6.224\,68-12&1.302\,46-08&1.937\,59-07\\
 5&3.014\,07-07&7.967\,25-08&4.758\,87-07&1.685\,06-08&1.925\,14-07&5.569\,65-07\\
  &1.613\,88-07&2.709\,90-08&2.626\,29-07&3.585\,87-09&6.915\,50-08&3.170\,65-07\\
  &6.941\,20-08&1.811\,97-09&1.146\,16-07&1.565\,94-10&6.048\,80-09&1.402\,78-07\\
 6&3.207\,99-07&9.285\,79-08&5.009\,67-07&2.109\,87-08&2.217\,19-07&5.794\,49-07\\
  &1.567\,37-07&2.666\,41-08&2.546\,75-07&3.571\,42-09&6.794\,44-08&3.069\,22-07\\
  &4.952\,07-08&5.182\,10-10&8.159\,26-08&4.325\,02-10&2.152\,92-09&9.958\,45-08\\
 7&3.294\,90-07&1.042\,47-07&5.081\,31-07&2.527\,53-08&2.459\,44-07&5.797\,94-07\\
  &1.520\,72-07&2.626\,02-08&2.466\,56-07&3.565\,77-09&6.680\,07-08&2.966\,42-07\\
  &3.456\,91-08&3.448\,24-11&5.676\,83-08&7.573\,73-10&3.815\,77-10&6.900\,63-08\\
 8&3.278\,82-07&1.133\,81-07&4.984\,70-07&2.918\,88-08&2.642\,39-07&5.598\,86-07\\
  &1.472\,13-07&2.585\,04-08&2.382\,86-07&3.563\,40-09&6.563\,11-08&2.858\,90-07\\
  &2.353\,32-08&6.950\,92-11&3.846\,88-08&1.078\,40-09&9.972\,57-12&4.650\,68-08\\
 9&3.166\,31-07&1.198\,77-07&4.735\,30-07&3.265\,26-08&2.758\,66-07&5.222\,05-07\\
  &1.420\,95-07&2.541\,62-08&2.294\,66-07&3.561\,15-09&6.438\,97-08&2.745\,59-07\\
  &1.557\,36-08&4.029\,37-10&2.530\,54-08&1.361\,71-09&4.833\,24-10&3.037\,65-08\\
10&2.967\,79-07&1.234\,84-07&4.354\,87-07&3.550\,39-08&2.804\,30-07&4.699\,78-07\\
  &1.367\,05-07&2.494\,77-08&2.201\,84-07&3.556\,97-09&6.305\,21-08&2.626\,41-07\\
  &9.984\,06-09&8.730\,29-10&1.609\,93-08&1.588\,31-09&1.388\,38-09&1.915\,21-08\\
11&2.697\,51-07&1.241\,05-07&3.871\,01-07&3.761\,71-08&2.779\,21-07&4.070\,73-07\\
  &1.310\,56-07&2.443\,93-08&2.104\,68-07&3.549\,43-09&6.160\,56-08&2.501\,85-07\\
  &6.174\,74-09&1.367\,15-09&9.861\,25-09&1.750\,47-09&2.431\,87-09&1.159\,85-08\\
12&2.372\,84-07&1.218\,08-07&3.315\,58-07&3.891\,19-08&2.687\,08-07&3.377\,66-07\\
  &1.251\,76-07&2.388\,83-08&2.003\,72-07&3.537\,51-09&6.004\,43-08&2.372\,66-07\\
  &3.664\,57-09&1.812\,72-09&5.781\,88-09&1.848\,36-09&3.419\,76-09&6.703\,66-09\\
13&2.013\,14-07&1.168\,10-07&2.722\,68-07&3.935\,61-08&2.534\,97-07&2.664\,61-07\\
  &1.191\,02-07&2.329\,38-08&1.899\,63-07&3.520\,50-09&5.836\,71-08&2.239\,77-07\\
  &2.072\,49-09&2.168\,45-09&3.220\,15-09&1.887\,28-09&4.236\,60-09&3.666\,07-09\\
14&1.638\,56-07&1.094\,48-07&2.126\,54-07&3.896\,32-08&2.332\,57-07&1.974\,23-07\\
  &1.128\,75-07&2.265\,65-08&1.793\,15-07&3.497\,94-09&5.657\,63-08&2.104\,17-07\\
  &1.106\,44-09&2.416\,35-09&1.685\,68-09&1.875\,42-09&4.826\,21-09&1.874\,47-09\\
15&1.268\,86-07&1.001\,55-07&1.559\,56-07&3.778\,72-08&2.091\,41-07&1.345\,41-07\\
  &1.065\,38-07&2.197\,77-08&1.685\,05-07&3.469\,59-09&5.467\,65-08&1.966\,90-07\\
  &5.503\,02-10&2.554\,61-09&8.171\,40-10&1.822\,20-09&5.174\,46-09&8.808\,62-10\\
16&9.223\,82-08&8.942\,07-08&1.050\,69-07&3.591\,44-08&1.823\,95-07&8.114\,32-08\\
  &1.001\,36-07&2.125\,98-08&1.576\,12-07&3.435\,36-09&5.267\,42-08&1.829\,00-07\\
  &2.501\,72-10&2.591\,95-09&3.588\,70-10&1.737\,09-09&5.294\,99-09&3.708\,12-10\\
17&6.152\,63-08&7.775\,80-08&6.243\,16-08&3.345\,47-08&1.542\,90-07&3.987\,69-08\\
  &9.371\,07-08&2.050\,57-08&1.467\,11-07&3.395\,33-09&5.057\,72-08&1.691\,46-07\\
  &1.009\,96-10&2.543\,02-09&1.380\,41-10&1.628\,96-09&5.217\,88-09&1.342\,71-10\\
18&3.608\,52-08&6.567\,76-08&2.994\,76-08&3.053\,21-08&1.260\,48-07&1.264\,35-08\\
  &8.730\,54-08&1.971\,86-08&1.358\,76-07&3.349\,67-09&4.839\,45-08&1.555\,25-07\\
  &3.456\,07-11&2.425\,16-09&4.397\,10-11&1.505\,59-09&4.981\,25-09&3.900\,66-11\\
19&1.694\,12-08&5.366\,17-08&8.957\,52-09&2.727\,70-08&9.879\,86-08&5.847\,00-10\\
  &8.095\,94-08&1.890\,20-08&1.251\,77-07&3.298\,61-09&4.613\,53-08&1.421\,29-07\\
  &9.236\,20-12&2.256\,10-09&1.046\,55-11&1.373\,54-09&4.625\,31-09&7.945\,85-12\\
20&4.801\,86-09&4.214\,82-08&2.402\,42-10&2.381\,90-08&7.353\,80-08&4.109\,05-09\\
  &7.471\,00-08&1.805\,97-08&1.146\,77-07&3.242\,44-09&4.380\,95-08&1.290\,40-07\\
  &1.634\,20-12&2.052\,46-09&1.491\,48-12&1.238\,14-09&4.188\,47-09&8.263\,64-13\\
21&6.367\,84-11&3.151\,84-08&4.042\,78-09&2.028\,15-08&5.111\,21-08&2.294\,84-08\\
  &6.859\,13-08&1.719\,52-08&1.044\,37-07&3.181\,49-09&4.142\,69-08&1.163\,35-07\\
  &1.257\,21-13&1.828\,90-09&6.692\,18-14&1.103\,56-09&3.705\,00-09&1.417\,34-14\\
22&2.833\,83-09&2.209\,23-08&2.012\,90-08&1.677\,84-08&3.220\,67-08&5.623\,30-08\\
  &6.263\,42-08&1.631\,23-08&9.450\,89-08&3.116\,02-09&3.899\,71-08&1.040\,85-07\\\cline{7-7}
  &6.603\,90-16&1.597\,74-09&8.018\,89-18&9.729\,19-10&3.203\,89-09&{\it 6.866\,45-18}\\
23&1.296\,07-08&1.412\,66-08&4.783\,98-08&1.341\,17-08&1.735\,01-08&1.025\,75-07\\
  &5.686\,63-08&1.541\,45-08&8.494\,09-08&3.046\,28-09&3.652\,98-08&9.234\,86-08\\\cline{2-2}\cline{4-4}
  &{\it 2.025\,61-16}&1.368\,88-09&{\it 7.058\,38-15}&8.484\,91-10&2.708\,50-09&{\it 7.298\,79-14}\\
24&3.006\,99-08&7.818\,20-09&8.615\,82-08&1.027\,11-08&6.922\,56-09&1.601\,54-07\\
  &5.131\,20-08&1.450\,52-08&7.577\,44-08&2.972\,44-09&3.403\,39-08&8.118\,05-08\\
  &{\it 6.639\,13-14}&1.149\,88-09&{\it 3.825\,31-13}&7.318\,12-10&2.236\,72-09&{\it 1.475\,85-12}\\
25&5.360\,26-08&3.308\,86-09&1.337\,74-07&7.434\,48-09&1.170\,61-09&2.267\,95-07\\
  &4.599\,22-08&1.358\,74-08&6.704\,54-08&2.894\,56-09&3.151\,83-08&7.062\,60-08\\
  &{\it 8.591\,82-13}&9.462\,12-10&{\it 3.286\,54-12}&6.238\,49-10&1.801\,45-09&{\it 9.232\,71-12}\\
26&8.285\,30-08&6.929\,09-10&1.891\,45-07&4.969\,60-09&2.243\,78-10&3.000\,34-07\\
  &4.092\,49-08&1.266\,43-08&5.878\,42-08&2.812\,55-09&2.899\,09-08&6.072\,32-08\\
  &{\it 4.483\,42-12}&7.615\,46-10&{\it 1.433\,78-11}&5.251\,26-10&1.411\,26-09&{\it 3.453\,91-11}\\
27&1.170\,03-07&2.547\,68-11&2.505\,47-07&2.936\,77-09&4.115\,31-09&3.771\,70-07\\
  &3.612\,48-08&1.173\,84-08&5.101\,58-08&2.726\,10-09&2.645\,96-08&5.150\,34-08\\
  &{\it 1.515\,17-11}&5.980\,33-10&{\it 4.393\,05-11}&4.358\,31-10&1.071\,06-09&{\it 9.690\,71-11}\\
28&1.551\,55-07&1.331\,45-09&3.161\,14-07&1.392\,31-09&1.279\,22-08&4.552\,88-07\\
  &3.160\,38-08&1.081\,22-08&4.376\,02-08&2.634\,66-09&2.393\,13-08&4.299\,11-08\\
  &{\it 3.973\,15-11}&4.566\,04-10&{\it 1.085\,29-10}&3.559\,08-10&7.827\,86-10&{\it 2.267\,22-10}\\
29&1.963\,55-07&4.614\,58-09&3.838\,68-07&3.931\,96-10&2.613\,24-08&5.312\,52-07\\
  &2.737\,13-08&9.887\,88-09&3.703\,24-08&2.537\,30-09&2.141\,25-08&3.520\,44-08\\
  &{\it 8.836\,10-11}&3.372\,29-10&{\it 2.323\,95-10}&2.851\,33-10&5.460\,45-10&{\it 4.691\,66-10}\\
30&2.396\,14-07&9.866\,27-09&4.517\,18-07&2.792\,67-12&4.394\,41-08&6.016\,44-07\\
  &2.343\,40-08&8.967\,20-09&3.084\,36-08&2.432\,53-09&1.890\,86-08&2.815\,61-08\\
  &{\it 1.753\,59-10}&2.391\,47-10&{\it 4.503\,88-10}&2.231\,64-10&3.586\,27-10&{\it 8.911\,70-10}\\
31&2.839\,17-07&1.707\,35-08&5.174\,45-07&2.981\,56-10&6.594\,74-08&6.626\,16-07\\
  &1.979\,66-08&8.051\,67-09&2.520\,07-08&2.318\,03-09&1.642\,37-08&2.185\,36-08\\
  &{\it 3.207\,59-10}&1.610\,56-10&{\it 8.129\,59-10}&1.695\,91-10&2.169\,60-10&{\it 1.594\,77-09}\\
32&3.282\,30-07&2.622\,48-08&5.786\,27-07&1.378\,71-09&9.170\,79-08&7.095\,59-07\\
  &1.646\,15-08&7.142\,29-09&2.010\,72-08&2.190\,00-09&1.395\,93-08&1.629\,95-08\\
  &{\it 5.530\,23-10}&1.012\,66-10&{\it 1.395\,55-09}&1.239\,65-10&1.164\,28-10&{\it 2.744\,97-09}\\
33&3.714\,82-07&3.730\,98-08&6.324\,92-07&3.373\,48-09&1.204\,39-07&7.363\,38-07\\
  &1.342\,94-08&6.239\,26-09&1.556\,31-08&2.041\,80-09&1.151\,08-08&1.149\,16-08\\
  &{\it 9.140\,72-10}&5.780\,87-11&{\it 2.317\,44-09}&8.581\,12-11&5.155\,74-11&{\it 4.635\,39-09}\\
34&4.125\,30-07&5.030\,24-08&6.755\,69-07&6.427\,78-09&1.503\,43-07&7.331\,27-07\\
  &1.069\,95-08&5.341\,33-09&1.156\,57-08&1.859\,72-09&9.057\,10-09&7.421\,17-09\\
  &{\it 1.469\,02-09}&2.850\,59-11&{\it 3.783\,25-09}&5.463\,02-11&1.598\,07-11&{\it 7.879\,45-09}\\
35&4.500\,71-07      &6.508\,15-08&7.028\,51-07      &1.05\hspace{0.6cm}-08&1.75\hspace{0.6cm}-07&6.76\hspace{0.6cm}-07\\
  &8.269\,35-09      &4.443\,93-09&8.108\,79-09      &1.60\hspace{0.6cm}-09&6.49\hspace{0.6cm}-09&4.05\hspace{0.6cm}-09\\
  &{\it 2.326\,54-09}&1.097\,77-11&{\it 6.188\,39-09}&2.97\hspace{0.6cm}-11&2.05\hspace{0.6cm}-12&{\it 1.4\hspace{0.75cm}-08}\\
36&4.824\,27-07      &8.108\,3\,\,\,-08    &7.051\,4\,\,\,-07      &&&\\
  &6.135\,07-09      &3.532\,6\,\,\,-09    &5.179\,8\,\,\,-09      &&&\\
  &{\it 3.686\,47-09}&2.566\,2\,\,\,-12    &{\it 1.046\,7\,\,\,-08}&&&\\
37&5.069\,63-07      &9.4\hspace{0.75cm}-08&6.5\hspace{0.75cm}-07  &&&\\
  &4.291\,06-09      &2.5\hspace{0.75cm}-09&2.7\hspace{0.75cm}-09  &&&\\
  &{\it 5.971\,46-09}&1.3\hspace{0.75cm}-13&                       &&&\\
38&5.179\hspace{0.4cm}-07      &&&&&\\
  &2.727\hspace{0.4cm}-09      &&&&&\\
  &{\it 1.033\hspace{0.4cm}-08}&&&&&\\
\end{longtable}
%}
\end{center}

\begin{figure}[hbt]
\setlength{\unitlength}{1mm}
\begin{picture}(140,50) (-20,10) 
\put(0,0){\mbox{\scalebox{1.2}{\includegraphics{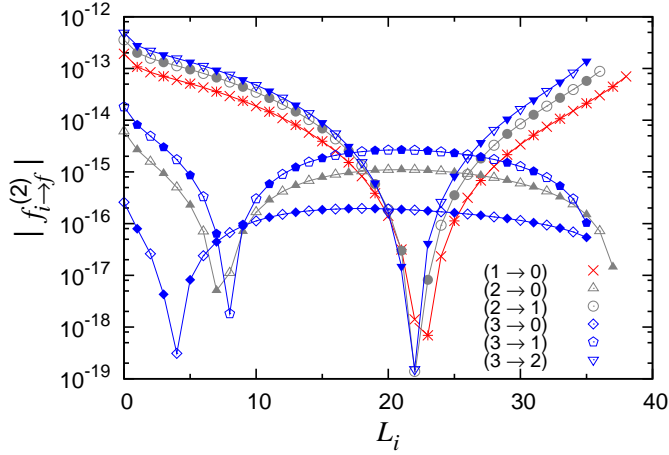}}}}
\end{picture} \\
\caption{Oscillator strengths for $L_f = L_i + 2$ transitions.}
\label{fig:2}
\end{figure}
\begin{figure}[hbt]
\setlength{\unitlength}{1mm}
\begin{picture}(140,55) (-20,10) 
\put(0,0){\mbox{\scalebox{1.2}{\includegraphics{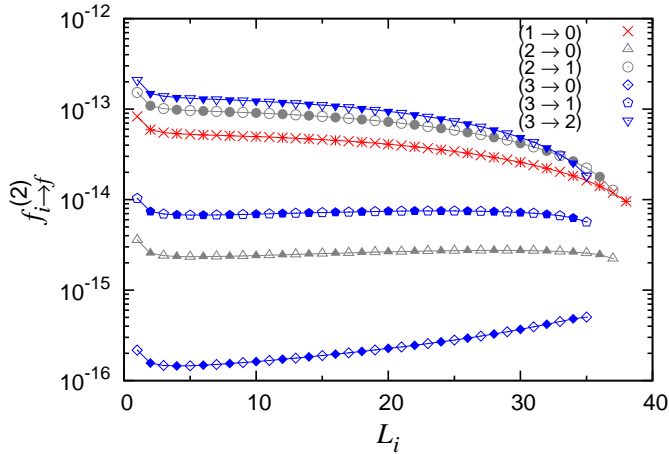}}}}
\end{picture} \\
\caption{Oscillator strengths for $L_f = L_i$ transitions.}
\label{fig:3}
\end{figure}
\begin{figure}[hbt]
\setlength{\unitlength}{1mm}
\begin{picture}(140,55) (-20,10) 
\put(0,0){\mbox{\scalebox{1.2}{\includegraphics{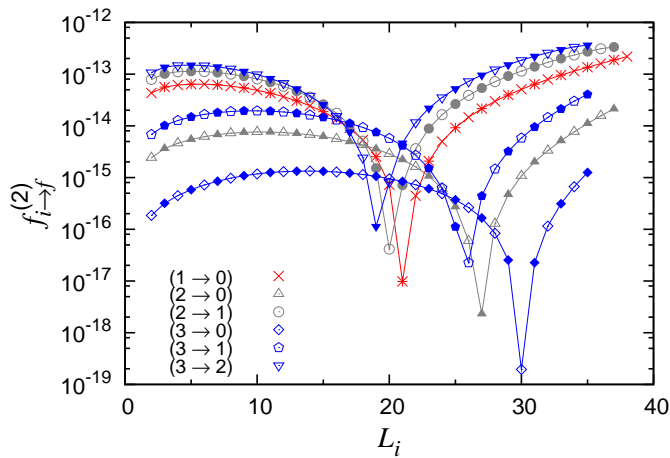}}}}
\end{picture} \\
\caption{Oscillator strengths for $L_f = L_i - 2$ transitions.}
\label{fig:4}
\end{figure}

Oscillator strengths are depicted in Figs.~\ref{fig:2}, \ref{fig:3} and \ref{fig:4}. 
For the transitions with $\Delta L = L_i - L_f = -2$ displayed in Fig.~\ref{fig:2}, 
they present a strong variation along the band. 
They also vary strongly with $\Delta v = v_i - v_f$. 
The $\Delta v = 1$ transitions present a deep minimum around $L_i = 23$ 
due to the change of sign of the energy difference (see Fig.~\ref{fig:1}). 
Beyond that value, the initial state is lower than the final $v_f < v_i$ state and 
the strengths are negative. 
Otherwise the strengths slowly increase with the vibrational excitation. 
The $\Delta v = 2$ strengths are smaller by more than an order of magnitude. 
The minimum occurring around $L_i = 8$ is here due to a change of sign of the matrix element 
appearing in \eref{2.3}. 
The $\Delta v = 3$ strengths are smaller by more than an order of magnitude 
than the $\Delta v = 2$ strengths. 
Except near the minimum at $L_i = 4$ also due to a change of sign of the matrix element, 
they are rather constant along the band. 

The $\Delta L = 0$ oscillator strengths presented in Fig.~\ref{fig:3} do not vary much along the bands 
as expected from the similar vibrational structures of the initial and final states. 
They slowly decrease for $\Delta v = 1$ and slowly increase for $\Delta v = 3$. 
They are remarkably flat for $\Delta v = 2$. 
Except near the end of the band, for given $\Delta v$, they increase with $v$. 

The $\Delta L = +2$ strengths presented in Fig.~\ref{fig:4} behave similarly 
to the $\Delta L = -2$ strengths. 
Minima take place around $L_i = 20$ for $\Delta v = 1$, $L_i = 26$ for $\Delta v = 2$ 
and $L_i = 30$ for $\Delta v = 3$. 
These minima thus occur now at increasing $L_i$ values with increasing $\Delta v$ 
and are all due to a change of sign of the matrix element. 

\begin{figure}[hbt]
\setlength{\unitlength}{1mm}
\begin{picture}(140,60) (-20,7) 
\put(0,0){\mbox{\scalebox{1.2}{\includegraphics{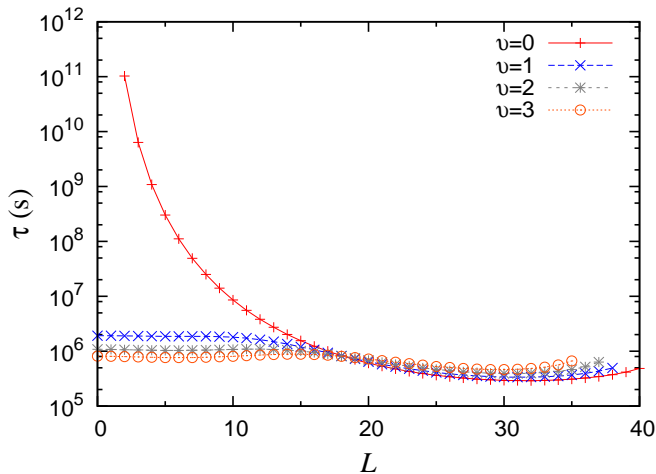}}}}
\end{picture} \\
\caption{Lifetimes $\tau$ in the first four rotational bands ($v = 0-3$).}
\label{fig:5}
\end{figure}

Lifetimes are defined as 
\beq
\tau = \bigg( \sum_{E_f < E_i} W_{i \rightarrow f} \bigg)^{-1}.
\eeqn{6.1}
For the $L = 0$ vibrational states, they have been calculated in \cite{PHB79}. 
The values $1.92 \times 10^6$ s, $1.08 \times 10^6$ s and $0.813 \times 10^6$ s 
for the first, second and third $L = 0$ excited vibrational states 
agree respectively well with our values $1.917\,518 \times 10^6$ s, $1.081\,130 \times 10^6$ s 
and $0.812\,900 \times 10^6$ s obtained from Table \ref{tab:4}. 
Lifetimes for the calculated states are displayed in Fig.~\ref{fig:5}. 
For the $(1^-,0)$ state, the lifetime is infinite within the present description. 
The large difference between the lifetimes of the $(2^+,0)$ state and, for example, 
the $(0^+,1)$ state is essentially due to the factor $(E_i - E_f)^5$ in expression \eref{2.8} 
of the E2 transition probability. 
Indeed the vibrational energy difference in the transition from the $(0^+,1)$ state 
is about 10 times larges than the rotational energy difference in the transition from 
the $(2^+,0)$ state which leads to an order of magnitude of $10^5$ for the ratio 
of the lifetimes. 

Except for the ground-state rotational band, where the lifetimes decrease rather fast from 
the very high values obtained at low orbital momenta (about 3300 years for $L = 2$), 
the lifetimes do not depend much on $L$. 
They are rather constant up to about $L = 15$. 
Then they slowly decrease till about $L = 30$ before starting increasing again slowly. 
The decrease is less than a factor of five for $v=1$, three for $v=2$ and two for $v=3$. 
For $v = 1$, the lifetimes are of the order of twenty-two days at small $L$ 
and of four days near $L = 30$. 
When $v$ increases from 1 to 3, they decrease progressively for $L \le 18$ and 
increase progressively for $L > 18$. 
\section{Conclusion}
\label{sec:conc}
In this paper, by accurately solving a three-body Schr\"odinger equation with Coulomb potentials, 
we have calculated the energies and wave functions of up to four 
of the lowest vibrational bound or quasibound states of the hydrogen molecular ion 
from $L = 0$ to $40$. 
The calculation is performed in perimetric coordinates with the Lagrange-mesh method. 
For low orbital momenta, a comparison with more accurate calculations in the literature 
and a cross comparison with the extensive results of Moss \cite{Mo93} 
show that the energies with 40 mesh points for the $x$ and $y$ coordinates 
and 20 mesh points for the $z$ coordinate have an accuracy of about 13 digits 
for the lowest vibrational state and a slowly decreasing accuracy with vibrational excitation 
providing still at least 9 digits for the third excited vibrational state. 
These accuracies are maintained along the whole rotational bands. 

With the corresponding wave functions, a simple calculation using the associated 
Gauss-Laguerre quadrature provides the quadrupole strengths 
and transition probabilities per time unit over the whole rotational bands. 
Tests with increasing numbers of mesh points and various truncations on $K$ show 
that the accuracy on this probabilities should reach six significant figures. 
The $K=0$ approximation leads to an error smaller than 0.3 \%. 
It slightly differs from the Born-Oppenheimer approximation by the fact 
that the proton mass is taken into account here. 
For the calculated states, we display tables extending the results 
and improving the accuracy of \cite{PDP83}. 
Although the displayed accuracy may exceed what is needed in applications, 
the results can also serve as a benchmark for testing approximate wave functions of H$_2^+$. 
Except for the states in the ground-state rotational band that can only decay slowly 
to the $L_f = L_i - 2$ previous state, 
all calculated lifetimes have an order of magnitude around $10^6$ s. 
\ack
We thank Michel Hesse for information about the codes. 
This text presents research results of 
the interuniversity attraction pole programme P6/23 initiated by the Belgian-state 
Federal Services for Scientific, Technical and Cultural Affairs. 
H.O.P. thanks CONACyT (M\'exico) for a postdoctoral grant.

\section*{References}
%\bibliographystyle{iopart-num}
%\bibliography{h2p}

%
\end{document}